\documentclass[journal]{IEEEtran}

\ifCLASSINFOpdf
\else
   \usepackage[dvips]{graphicx}
\fi
\usepackage{url}

\usepackage{graphicx}
\usepackage{cite}
\usepackage{amsmath,amssymb,amsfonts,adjustbox,multirow}
\newcommand{\R}{\mathbb{R}}
\usepackage{algorithmic}
\usepackage{array}
\newcolumntype{P}[1]{>{\centering\arraybackslash}p{#1}}
\newcolumntype{M}[1]{>{\centering\arraybackslash}m{#1}}
\usepackage{enumitem}
\setlist{leftmargin=*}

\usepackage{subfigure}
\usepackage{helvet}
\usepackage[table,xcdraw]{xcolor}
\usepackage{graphicx}
\usepackage{textcomp}

\begin{document}

\title{Multi-Label ECG Classification using Temporal Convolutional Neural Network}

\author{Eedara Prabhakararao and  Samarendra Dandapat, \IEEEmembership{Member, IEEE}

\thanks{Eedara Prabhakararao and Samarendra Dandapat are with the Department
of Electronics and Electrical Engineering, Indian Institute of Technology
Guwahati, India-781039 (e-mails: e.prabha, samaren@iitg.ac.in).}}

\markboth{Journal of \LaTeX\ Class Files, Vol. 14, No. 8, August 2015}
{Shell \MakeLowercase{\textit{et al.}}: Bare Demo of IEEEtran.cls for IEEE Journals}
\maketitle

\begin{abstract}
Automated analysis of 12-lead electrocardiogram (ECG) plays a crucial role in the early screening and management of cardiovascular diseases (CVDs). In practice, it is common to see multiple co-occurring cardiac disorders, i.e., multi-label or multimorbidity in patients with CVDs, which increases the risk for mortality. Most current research focuses on the single-label ECG classification, i.e., each ECG record corresponds to one cardiac disorder, ignoring ECG records with multi-label phenomenon. In this paper, we propose an ensemble of attention-based temporal convolutional neural network (ATCNN) models for the multi-label classification of 12-lead ECG records. Specifically, a set of ATCNN-based single-lead binary classifiers are trained one for each cardiac disorder, and the predictions from these classifiers with simple thresholding generate the final multi-label decisions. The ATCNN contains a stack of TCNN layers followed by temporal and spatial attention layers. The TCNN layers operate at different dilation rates to represent the multi-scaled pathological ECG features dynamics, and attention layers emphasize/reduce the diagnostically relevant/redundant 12-lead ECG information. The proposed framework is evaluated on the PTBXL-2020 dataset and achieved an average F1-score of 76.51$\%$. Moreover, the spatial and temporal attention weights visualization provides the optimal ECG-lead subset selection for each disease and model interpretability results, respectively. The improved performance and interpretability of the proposed approach demonstrate its ability to screen multimorbidity patients and help clinicians initiate timely treatment.
\end{abstract}


\begin{IEEEkeywords}
Arrhythmia, ECG, multi-label classification, deep learning, model interpretability, optimal leads selection.
\end{IEEEkeywords}

\IEEEpeerreviewmaketitle

\section{Introduction}
\label{sec:introduction}
\IEEEPARstart{C}{ardiovascular} diseases (CVDs) are the leading cause of mortality worldwide, taking over 18.6 million lives in 2019 \cite{Roth2020}. Most CVD-related deaths are associated with multimorbidity, i.e., the co-occurrence of two or more acute cardiac disorders such as myocardial infarction and ischemia, hypertrophy, bundle branch blocks, atrial and ventricular fibrillation in the same individual \cite{Buddeke2017}. The presence of multiple cardiac disorders in an individual could potentially lead to poorer functional status, lower quality of life (QoL), and increased risk of mortality \cite{Buddeke2017,Rahimi2018}. Therefore, early detection of multimorbidity patients is crucial to prevent adverse health outcomes and optimize patient-centered care delivery \cite{Rahimi2018}.

The standard 12-lead electrocardiogram (ECG), including lead I, II, III, aVR, aVL, aVF, and V1 to V6,  represents the heart's electrical activity from different spatial angles and is a widely used non-invasive diagnostic tool to screen patients with various cardiac disorders \cite{Begg2016}. Clinicians detect cardiac disorders by manually inspecting the changes in ECG characteristics such as P-wave, QRS-complex, ST-segment, and T-wave across the 12-lead ECG \cite{Begg2016}. However, even for an experienced clinician, accurately interpreting the 12-lead ECG for a patient with co-occurring cardiac disorders is challenging \cite{Zhu2020}. Consequently, there is an urgent unmet need for automated ECG interpretation methods for detecting co-occurring cardiac disorders. Most current studies focus on single-label ECG classification \cite{Acharya2017,Li2016,Kiranyaz2015,Ebrahimi2020,Rajpurkar2019,Qin2021,Hou2020,Meng2021,Wang2020,Yao2020,Zhang2021,Prabhakararao2021}, i.e., one ECG record corresponds to one cardiac disorder or label. However, considering the increased prevalence (more than 50$\%$) of multimorbidity in CVD patients and its association with a higher risk of mortality \cite{Buddeke2017,Rahimi2018}, the research towards the multi-label ECG classification, i.e., one ECG record corresponds to multiple concurrent cardiac disorders at the same time has been gaining popularity. 

In the literature, although few machine learning (ML) \cite{Li2020,Sun2020,Li2021} and deep learning (DL) \cite{Zhu2020,Cheng2020,Jia2019,Yoo2021,Ganeshkumar2021,Jin2021} based approaches have been investigated for multi-label ECG classification, two critical challenges still need to be addressed. (i) Existing approaches employed algorithm adaptation methods, where they directly train the model with multi-labeled ECG data assuming that the training dataset consists of all the possible label combinations. For example, training a DL model for a ten ECG labels problem requires data for $2^{10}$ number of label combinations, making the model learning cumbersome and inefficient as the real-world ECG data does not have many label combinations. In addition, the ECG data for different disease labels is highly imbalanced; thus, choosing proper thresholds for each label is a challenging task to achieve reliable multi-label ECG classification performance. For example, if the number of thresholds is 10 (for ten labels), and 15 values are assessed for each threshold, the number of threshold combinations would be $15^{10}$, thus making threshold selection infeasible. (ii) The different leads of the 12-lead ECG exhibit distinct temporal variations in the ECG characteristics that appear at different shapes and morphology (multi-scale) associated with specific types of cardiac disorder \cite{Fuster2013}. For example, the right bundle branch block (RBBB) is diagnosed by the abnormal QRS pattern at lead V1/V2 and the slur S-wave pattern at leads I/aVL, whereas the large R-wave at leads V5/V6 detect the left ventricular hypertrophy (LVH) \cite{Fuster2013}. The myocardial infarction is characterized by the ST-segment elevation at the leads facing the injured cardiac muscle \cite{Fuster2013}. In addition, concurrent cardiac disorders result in highly irregular ECG features, thus increasing the difficulty of multi-label ECG classification \cite{Zhu2020}. Notably, the 12-lead ECG shows implicit diagnostic redundancy, which often imposes systematic overfitting on the DL model causing poor generalization. 

This article presents a new attention-based temporal convolutional neural network (ATCNN) to address the above challenges. The major contributions in this paper are as follows: 
\begin{enumerate}
    \item We propose a generic deep neural network architecture for multi-label ECG classification using the problem transformation approach. Using this approach, we transform the multi-label problem into an ensemble of single-label binary classifiers trained one for each cardiac disorder type. The diagnosis decisions from the binary classifiers are used to detect multiple co-occurring cardiac disorders appearing in the same ECG record. The model design is more efficient than the algorithm adaptation methods for multi-label ECG analysis and can be extended to any additional disorders with relatively minimal effort.  
    \item We design each binary classifier using the attention-based temporal convolutional neural network (ATCNN). The ATCNN consists of a stack of temporal convolutional layers followed by temporal and spatial attention modules. The temporal layers are designed using dilated convolution filters with different sizes or receptive fields to extract the temporal dynamics of multi-scaled ECG features effectively. In addition, the temporal and spatial attention modules exploit the 12-lead ECG's within-lead and across-leads disease-specific diagnostic information, respectively, and enhance the multi-lead feature extraction pipeline.
    \item For the first time in the literature, we investigate the spatial attention weights-based optimal subset ECG leads selection for each disease category. We demonstrate that the ATCNN model automatically identifies the optimal ECG-lead subset, alleviates the 12-lead ECG redundancy, and improves model generalization. In addition, we show that the proposed multi-label ECG classification framework can identify low- and high-risk patients assessed in terms of the number of co-occurring cardiac disorders.    
\end{enumerate} 
The rest of this paper is organized as follows: Section II presents the related works on multi-label ECG classification. Section III describes the proposed method. Section IV presents the experimental results. Section V gives the discussion. Finally, Section VI concludes the paper.

\section{Related works}
In the literature, few works have been reported for the multi-label ECG classification. Traditional methods \cite{Li2020,Sun2020,Li2021} use domain expert knowledge to extract various handcrafted features, followed by ML classifiers to detect multi-label ECG abnormalities. Sun \emph{et al.} \cite{Sun2020} proposed an ensemble classifier for multi-label ECG classification. The method employed 169 handcrafted features, including morphological, signal quality, non-linear, time-, and frequency-domain features to train an ensemble of several ML-based multi-label classifiers to classify the multi-label ECG recordings. Recently, Li \emph{et al.} \cite{Li2020,Li2021} presented an ML-based multi-objective optimization model that exploits the disease correlations to classify the multi-label ECG signals. The method used 117 handcrafted features followed by feature selection and classification. The performance of these traditional methods highly depends on the robustness of detecting ECG fiducial points. Moreover, extracting reliable features from the ECG signals of patients with concurrent cardiac disorders is challenging and requires exceptional domain expertise.

During the past two years, DL-based models have demonstrated impressive performance for the multi-label ECG classification when trained with sufficient data \cite{Cheng2020,Jia2019,Yoo2021,Ganeshkumar2021,Jin2021}. The DL models can automatically learn useful features from the ECG data, thereby alleviating the need for manual feature extraction. Cheng \emph{et al.} \cite{Cheng2020} proposed a residual convolutional neural network (ResCNN) model for detecting multi-label ECG arrhythmias. An ensemble of sequence generation and multi-task modules is used in \cite{Jia2019} for effective multi-label rhythm abnormalities classification. A dual-attention DNN model integrating the advantages of CNNs, bidirectional long short-term memory networks (BiLSTM), and attention modules have been investigated in \cite{Jin2021} for an interpretable multi-label ECG classification. An explainable DL model combing CNNs and Grad-CAM has been proposed in \cite{Ganeshkumar2021} to analyze multi-label ECG abnormalities. In \cite{Yoo2021}, a residual squeeze and excitation block (SE-ResNet) followed by a random k-labelset-based soft voting is investigated to improve the multi-label ECG classification. Most of the above approaches employed the algorithm adaption methods for multi-label ECG classification. However, this paper investigates the problem transformation method for effective multi-label ECG classification.

\begin{figure*}[ht!]
\centerline{\includegraphics[height=1.8in,width=6.8in]{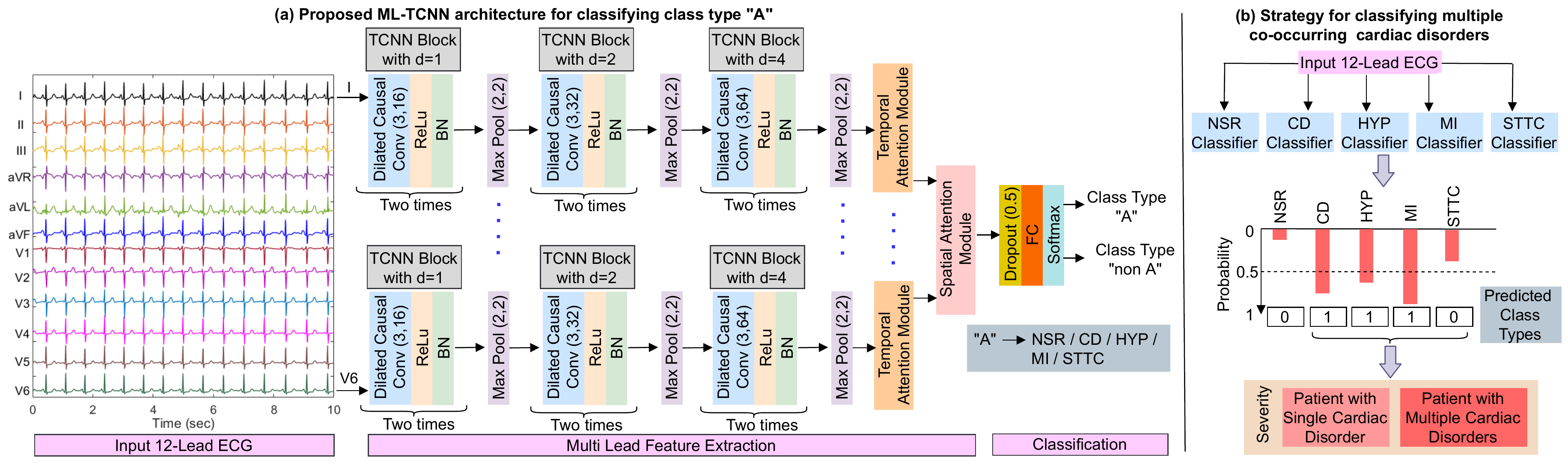}}
\caption{(a) Proposed ATCNN model architecture for single-label binary classification using 12-lead ECG signals. (b) Multi-label ECG classification strategy.}
\label{BD01}
\end{figure*}

\section{Proposed Methodology}
\subsection{Problem Formulation}
Multi-label ECG abnormalities detection belongs to a time series classification problem. Given a set of 12-lead ECG recordings and their corresponding co-occurring disease labels, the goal is to predict multiple co-occurring cardiac disorders present in each ECG record. Because the co-occurring disorders result in irregular ECG features across the 12-leads, their accurate classification is challenging. Moreover, the existing algorithm adaption methods are inefficient for the multi-label ECG analysis. Therefore, in this paper, we propose a problem transformation-based DL framework for multi-label classification of 12-lead ECG into four cardiac disorders such as conduction disturbances (CD), hypertrophy (HYP), myocardial infarction (MI), and ST/T changes (STTC). Specifically, we transform the complex multi-label classification problem into an ensemble of single-label binary classifiers trained one for each cardiac disorder type using the attention-based temporal convolutional neural network (ATCNN).

\subsection{The Proposed Multi-Label ECG Classification Framework}
Fig. \ref{BD01}(b) presents an overview of the proposed multi-label ECG classification approach. As can be seen, first, the incoming 12-lead ECG input is fed to an ensemble of single-label binary classifiers trained independently, one for each cardiac disorder type. Second, the diagnosis predictions from the binary classifiers are thresholded to identify the multi-label ECG abnormalities. Fig. \ref{BD01}(a) illustrates the proposed ATCNN model architecture used to design each disease-specific binary classifier (disease exists or not). The ATCNN  contains two modules. The details of the modules are described below. 
\subsubsection{Multi-Lead Feature Extraction}
\begin{figure}[ht!]
\centerline{\includegraphics[height=1in,width=2.6in]{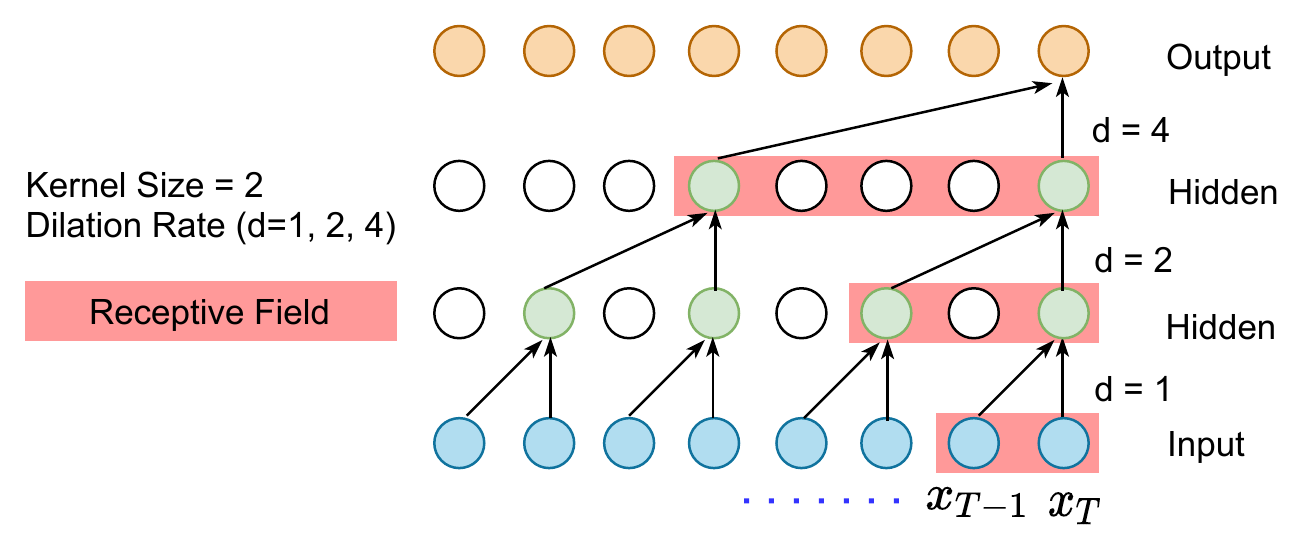}}
\caption{Example of dilated and causal convolutions in TCNN layers with filter size K = 2 and dilation rate d = [1, 2, 4].}
\label{BD02}
\end{figure}

This module automatically learns the informative features from the input 12-lead ECG using a stack of temporal CNN (TCNN) layers followed by temporal and spatial attention layers (Fig. \ref{BD01}(a)). The different leads of the 12-lead ECG exhibit distinct temporal variations in the ECG features that appear at different shapes and morphology. These lead-specific dynamics of multi-scaled ECG features are encoded using a stack of TCNN layers. Each TCNN layer operates at different receptive fields to effectively represents the multi-scaled disease variabilities (Fig. \ref{BD01}(a)). The varied receptive field size is attained using dilated and causal convolutional filters that operate at different dilation rates ($d$). Mathematically, for a 1D sequence input $\textbf{x} \in \R^T$ and a filter $\textbf{w} \in \R^{K}$, the dilated and causal convolution operation $\mathcal{F}$ on elements $n$ of the sequence is defined as:  
\begin{equation}
\mathcal{F}[n]  = (\textbf{x} \ast_d \textbf{w})[n] = \sum_{i=0}^{K-1} \textbf{w}[i] \cdot  \textbf{x}[n-d \cdot i]
\end{equation}
where $T$ and $K$ denotes the sequence length and the filter size, respectively. $[n-d \cdot i]$ represents the direction of past. The dilated and causal filter is obtained by inserting holes between the kernel elements of $\textbf{w}$ based on the dilation rate $d$. The effective receptive field size $K'$ of the dilated and causal filter with a dilation rate $d$ can be computed as $K' = (K-1)d + 1$. Fig. \ref{BD02} shows TCNN layers with different dilation rates. In traditional CNNs, the receptive field size can only increase linearly with the layer depth. However,  in the dilated and causal convolutions, the dilation rate increases exponentially with each layer (Fig. \ref{BD02}). Therefore, although the number of parameters increases linearly, the effective receptive field size grows exponentially with the layer depth. Moreover, recent studies \cite{Bai2018} demonstrate that the TCNN layers with dilated and causal convolution filters outperform the recurrent neural networks (RNNs) on various sequence modeling tasks by handling the temporal dependencies effectively.

Let $\textbf{X}_{ecg} \in \R^{12 \times T}$, $\textbf{X}_{ecg} = [\textbf{x}^1, \textbf{x}^2,...,\textbf{x}^m,...,\textbf{x}^{12}]^\mathsf{T}$ denote the input 12-lead ECG sequence. Here, $\mathsf{T}$ is the transpose operator and $\textbf{x}^m$ denote the $m^{th}$ ECG lead data with length $T$. The input 12-lead ECG ($\textbf{X}_{ecg}$) is fed to the multi-lead feature extraction module to extract the informative features. Fig. \ref{BD01}(a) illustrates the network configuration of the proposed multi-lead feature extractor consisting of a stack of TCNN layers followed by temporal and spatial attention layers. First, each lead of the 12-lead ECG is fed to a stack of TCNN layers that operate at different dilation rates to extract the multi-scaled ECG variabilities (Fig. \ref{BD01}(a)). The output feature maps of the TCNN layers for the $m^{th}$ ECG lead ($\textbf{x}^m$) can be given as:
\begin{equation}
\textbf{F}_m  = f^{TCNN} _{d=4}(f^{TCNN} _{d=2}[f^{TCNN} _{d=1} (\textbf{x}^m)])
\end{equation}
where $f^{TCNN} _d$ represents a TCNN block consisting of two TCNN layers with dilation rate $d$ and a stack of such blocks maps the $m^{th}$ ECG lead ($\textbf{x}^m$) to a output feature maps $\textbf{F}_m \in \R^{Z \times L}$, $\textbf{F}_m = [f^i _1, f^i _2,...,f^i _L]$ with $i=1, 2,...,Z$. Here, $Z$ and $L$ represents the number of feature maps and their length, respectively. Next, the lead-specific feature maps $\textbf{F}_m$ are fed to the temporal attention layer. This layer reduces the within-lead feature redundancy and aggregates the informative temporal feature representations using the temporal attention weights $\pmb{\alpha}_m \in \R^{1 \times L}$ computed as follows:
\begin{equation}
\pmb{\alpha}_m  = Softmax[tanh(\textbf{w}_m \textbf{F}_m + \textbf{b}_m)]
\end{equation}
where $\textbf{w}_m \in \R^{1 \times Z}$ is the weight vector and $\textbf{b}_m \in \R^{1 \times L}$ is the bias vector. The temporal attention weights ($\pmb{\alpha}_m$) of the $m^{th}$ ECG lead are used to combine the respective temporal feature representations computed as:
\begin{equation}
\textbf{u}_m  = \textbf{F}_m \pmb{\alpha}^\mathsf{T} _m
\end{equation}
where $\textbf{u}_m \in \R^{Z \times 1}$. The twelve lead-specific temporal summary vectors from the 12-leads are transformed into $\textbf{S} = [\textbf{u}_1, \textbf{u}_2,...,\textbf{u}_m,...,\textbf{u}_{12}]$, $\textbf{S} \in \R^{Z \times 12}$. Similar to the above formulation, the spatial attention layer takes the $\textbf{S}$ as an input and compute the final high-level representation ($\textbf{v} \in \R^{Z \times 1}$) used for classification. The spatial layer reduces the across-lead redundancy and aggregates the informative spatial feature representations using the spatial attention weights $\pmb{\beta} \in \R^{12 \times 1}$ computed as follows:  
\begin{align}
\begin{split}\label{eq:1}
\pmb{\beta}  = Softmax[tanh(\textbf{S}^\mathsf{T} \textbf{w}_s + \textbf{b}_s)] 
\end{split}\\
\begin{split}\label{eq:2}
\textbf{v}  = \textbf{S}   \pmb{\beta}
\end{split}
\end{align}
where $\textbf{w}_s$ and $\textbf{b}_s$ are the weight and bias vectors, respectively.
\subsubsection{Single-Label Binary Classification}
The high-level representation vector $\textbf{v}$ is fed to the fully-connected output layer with Softmax activation to compute the probability distribution of outputs for the input 12-lead ECG $\textbf{X}_{ecg}$ and is given as:
\begin{equation}
  \begin{bmatrix}
   p(A|\textbf{X}_{ecg}) \\
   1-p(A|\textbf{X}_{ecg})
   \end{bmatrix} = Softmax(\textbf{W}_o \textbf{v} + \textbf{b}_o)
\end{equation}
where $p(A|\textbf{X}_{ecg})$ and ($1-p(A|\textbf{X}_{ecg}$) represents the probability of $\textbf{X}_{ecg}$ belongs to class type $A$ and $nonA$, respectively and $A \in \{NSR, CD, HYP, MI, STTC\}$ (see Fig. \ref{BD01}(a)). 
\subsubsection{Loss Optimization}
The network parameters of the single-label binary classifier are trained or optimized using the binary cross entropy (BCE) loss function computed as:
\begin{equation}
BCE  = y log(p(A|\textbf{X}_{ecg})) + (1-y) log(1-p(A|\textbf{X}_{ecg}))
\end{equation}
where $y$ represents the true label of the input $\textbf{X}_{ecg}$, $y \in \{0,1\}$ with $1$ for class type $A$ and $0$ for class type $nonA$. 
\subsubsection{Multi-Label Classification}
After training a set of binary classifiers one for each class type, the multi-label ECG classification is performed as follows: first, the input $\textbf{X}_{ecg}$ is fed to the single-label binary classifies and obtained the probability of output class prediction as $[p(NSR|\textbf{X}_{ecg}); p(CD|\textbf{X}_{ecg}); p(HYP|\textbf{X}_{ecg});p(MI|\textbf{X}_{ecg});$ $ p(STTC|\textbf{X}_{ecg})]$. Finally, multiple co-occurring cardiac disorders (multi-label) in the same ECG record are determined with $0.5$ thresholding on the probability of predicting a specific class type (Fig. \ref{BD01}(b)). In addition, the proposed framework is employed to identify  low-  and  high-risk  patients  assessed  in  terms of the number of co-occurring cardiac disorders. 

\section{Experiment}
The proposed ATCNN model is implemented using Python and Keras framework. All the experiments are performed using an NVIDIA Tesla V100 GPU server. 
\subsection{Database Description}
In this study, we have used the PhysioNet PTBXL-2020 dataset \cite{Wagner2020} for model development and evaluation. The PTBXL-2020 is the largest publicly available multi-label 12-lead ECG waveform dataset comprising 21,837 records from 18,885 patients (females: 9,064; males: 9,821) recorded at the sampling rate of 100 Hz for 10 s duration. The database provides multi-label diagnosis annotations for each 12-lead ECG record in terms of NSR, CD, HYP, MI, and STTC. Also, the database provides the development and test partitions of the ECG records. The ECG records which do not have proper diagnosis annotations are excluded from the study. Table \ref{tab:Data} shows the distribution of ECG records across the five diagnostic labels and their concurrent or multi-label combinations for the development and test datasets. As can be seen, around 23$\%$ of records (4,798 out of 20,901) have more than one cardiac disorder at the same time; this emphasizes the importance of multi-label ECG classification. 

\subsection{Dataset for Training Binary Classifiers}
This study employs a problem transformation approach for multi-label ECG classification; thus, the multi-label dataset is divided into multiple sub-datasets representing each cardiac state. For example, the sub-dataset of MI cardiac state consists of MI and nonMI class ECG records. Here nonMI class consists of ECG records from all other cardiac states except MI. It is observed that the sub-datasets of each cardiac state (except NSR class) show an imbalanced ECG data distribution, i.e., the number of positive class samples is less than the negative class; thus, a random undersampling algorithm is employed (except for NSR) to mitigate class imbalance. For example, to train a binary classifier for the MI (MI vs. nonMI), the nonMI ECG recordings are selected randomly from the development dataset (see Table \ref{tab:Data}) until twice as the MI recordings. The selected sub-dataset is randomly divided into training and validation sets in the ratio of 8:2. It is to be noted that the random undersampling is applied only on the development dataset, and the test dataset (see Table I) is exclusively used for evaluating the performance of the proposed ATCNN model.

\subsection{Preprocessing}
The practical ECG signals are often corrupted by various noises, including muscle movement, power-line interference, and baseline wander, which can significantly alter the morphology of the ECG features \cite{Begg2016}. Therefore, to alleviate the possible effects of noise, each ECG-lead signal is processed using a Butterworth bandpass filter with a passband from 0.5 to 45 Hz. Moreover, the ECG signals amplitudes vary across different subjects; thus, each ECG-lead is z-score normalized to maintain zero mean and unit variance.

\begin{table}[t!] \scriptsize
\caption{Distribution of ECG recordings for the five diagnostic labels (NSR, CD, HYP, MI and STTC) and their combinations across the development and the test sets of the PTB-XL database} 
\centering 
\setlength{\tabcolsep}{2.3pt}
\renewcommand{\arraystretch}{1.2}
\begin{tabular}{ccc|ccc} \hline \hline
 \textbf{Label Type}  & \textbf{Development} & \textbf{Test}  &  \textbf{Concurrent Labels}   &  \textbf{Development} &  \textbf{Test} \\ \hline
 NSR    &  8093   & 902   & One-label   &  14465  &  1638 \\ 
 CD     &  4054   &  453  & Two-label   &  3335 & 349 \\
 HYP    &  2494   &  271  &  Three-label & 849 & 100 \\ 
 MI    &  4933   & 553   & Four-label  & 149  &  16 \\
 STTC    & 4704    & 519   & Five-label  & 0 &  0  \\ \hline
 \textbf{Total}   & 24278    & 2698   & \textbf{Total} & 18798  &  2103 \\ \hline        
\end{tabular}
\label{tab:Data}
\end{table}

\begin{figure*}[ht!]
\centerline{\includegraphics[height=1.4in,width=6.9in]{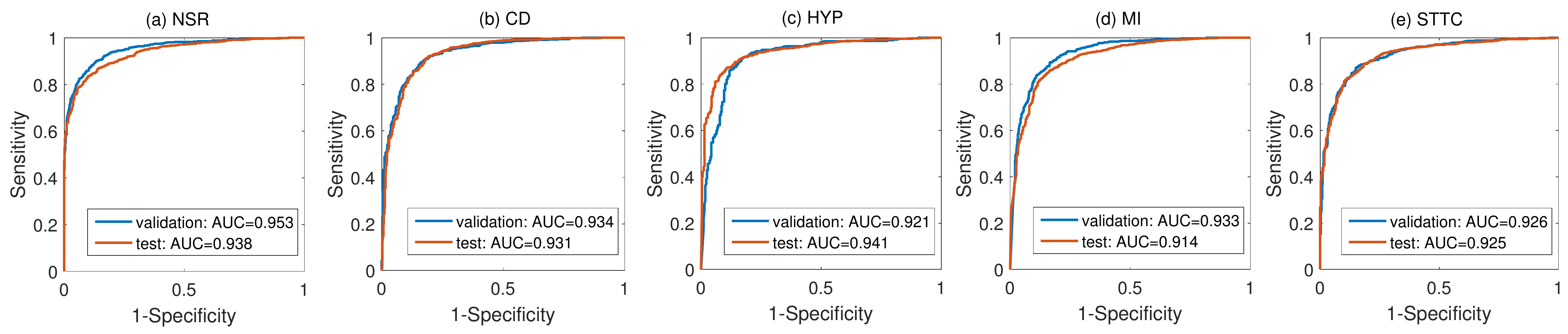}}
\caption{The receiver operating characteristic (ROC) curves show the prediction sensitivity of the proposed model for five cardiac conditions including (a) NSR, (b) CD, (c) HYP, (d) MI and (e) STTC on the validation and test datasets.}
\label{BD03}
\end{figure*}

\subsection{Model Configuration and Training Settings}
The proposed ATCNN configuration is described in Fig. \ref{BD01}(a). The two crucial hyperparameters, such as the number of TCNN blocks and the convolutional filter size, are optimized experimentally. We tested the number of TCNN blocks $\{1, 2, 3, 4\}$ with exponential increase in dilation rates and the filter sizes $\{1 \times 3, 1 \times 5, 1 \times 7\}$. The three TCNN blocks with filter size $1 \times 3$ provided high detection accuracy at less trainable parameters. The model parameters are trained end-to-end using a binary cross-entropy loss. The model is trained with an Adam optimizer with a learning rate of 0.001 on mini-batches of size 32 for 100 epochs. An early stopping criterion is employed to avoid overfitting, which stops the training when the performance does not improve over ten iterations. It is to be noted that the same ATCNN architecture is used for each cardiac condition (NSR/CD/HYP/MI/STTC), and their respective model parameters will be optimized during training.

\subsection{Evaluation Measures}
The classification performance of each cardiac condition is evaluated by recall (Re), precision (Pr), specificity (Sp), accuracy (Acc), and F1-score measures computed as:
\begin{align} 
\begin{split}\label{eq:11}
Recall  = \frac{TP}{TP+FN}
\end{split}\\
\begin{split}\label{eq:22}
Precision  = \frac{TP}{TP+FP}
\end{split}\\
\begin{split}\label{eq:33}
Specificity  = \frac{TN}{TN+FP}
\end{split}\\
\begin{split}\label{eq:44}
Accuracy  = \frac{TP+TN}{TP+TN+FP+FN}
\end{split}\\
\begin{split}\label{eq:55}
F1-score  = \frac{2 \times Precision \times Recall}{Precision + Recall}
\end{split}
\end{align}

where TP, TN, FP, FN represent the number of true positives, true negatives, false positives, and false negatives, respectively. To evaluate the exact match between the predicted and true multi-label vectors, an exact match measure is computed as:
\begin{equation}
Exact\,Match  =  \frac{1}{N} \sum_{l=1}^Q I (\textbf{y}_p = \textbf{y}_t)
\end{equation}
where $N$, $\textbf{I}(\cdot)$, $\textbf{y}_p \in \R^{c_l}$ and $\textbf{y}_t \in \R^{c_l}$ represent the number of test ECG records, indicator function, predicted and true multi-label vectors, respectively. $c_l$ represents the number of labels. The limitation of this measure is that it ignores partially correct predictions because they must all be matched.

\begin{table}[t!] \scriptsize
\caption{Proposed model performance on the test set} 
\centering 
\setlength{\tabcolsep}{2.8pt}
\renewcommand{\arraystretch}{1.2}
\begin{tabular}{c|cccc|ccccc} \hline \hline
 \textbf{Class}  & \textbf{TP} & \textbf{FN}  &  \textbf{FP}   &  \textbf{TN} &  \textbf{Re ($\%$)} &  \textbf{Pr ($\%$)} &  \textbf{Sp ($\%$)} &  \textbf{Acc ($\%$)} &  \textbf{F1-score ($\%$)} \\ \hline
 
 NSR    &  811   & 93   & 172   &  1027  &  89.71 &  82.50 &  85.65 & 87.40 & 86.01  \\ 
 
 CD    &  367   & 86   & 131   &  1519  &  81.01 &  73.69 &  92.06 & 89.68 & 77.18  \\
 
 HYP   &  230   & 41   & 190   &  1642  &  84.87 &  54.76 &  89.62 & 89.01 & 66.57  \\  
 
 MI    &  452   & 101   & 178   &  1372  &  81.73 &  71.74 & 88.56 & 86.73 & 76.41  \\ 
 
 STTC    &  426   & 93   & 171   &  1413  &  82.08 &  71.35 &  89.20 & 87.44 & 76.34  \\ \hline
 
 \multicolumn{5}{r|}{\textbf{Average}} &  \textbf{83.88} &  \textbf{70.80} &  \textbf{89.01} & \textbf{88.05} & \textbf{76.51}  \\ \hline 
 
  \multicolumn{5}{r|}{Multi-label exact match ($\%$)} &  \multicolumn{5}{c}{\textbf{66.38}}  \\ \hline 
 
\end{tabular}
\label{tab:results01}
\end{table}

\subsection{Results}
Table \ref{tab:results01} shows the class-wise and average classification performance of the proposed ATCNN model on the test dataset in terms of Re, Pr, Sp, Acc, and F1-score measures. As can be seen, the model achieved an average Re of 83.88$\%$, Pr of 70.80$\%$, Sp of 89.01$\%$, Acc of 88.05$\%$, F1-score of 76.51$\%$, respectively. Specifically, more than 81$\%$ of Re across all cardiac conditions shows that the proposed model has good sensitivity for detecting multiple cardiac disorders. Although the pathological ECG features that characterize CD, MI, and STTC  may be subtle and similar at certain ECG-leads, the model still achieved a promising F1-score of 77.18$\%$, 76.41$\%$ and 76.34$\%$, respectively. However, the under-representation of HYP samples in the training dataset slightly degrades its F1-score on the test dataset. Table \ref{tab:results01} also shows the TP, FN, FP, and TN metrics for each cardiac condition on the test dataset. As observed, most of the ECG records concerning the presence or absence (TP and TN) of various cardiac disorders are correctly classified with fewer misclassified (FP and FN) records. In addition, the proposed multi-label ECG classification method achieved a multi-label exact match measure of 66.38$\%$ on the test data (Table \ref{tab:results01}). The ROC curves plotted in Fig. \ref{BD03} highlight that the detection of all cardiac conditions had AUC higher than 0.91 on the test dataset, reflecting an impressive classification.

\begin{figure*}[ht!]
\centerline{\includegraphics[height=1.7in,width=6.6in]{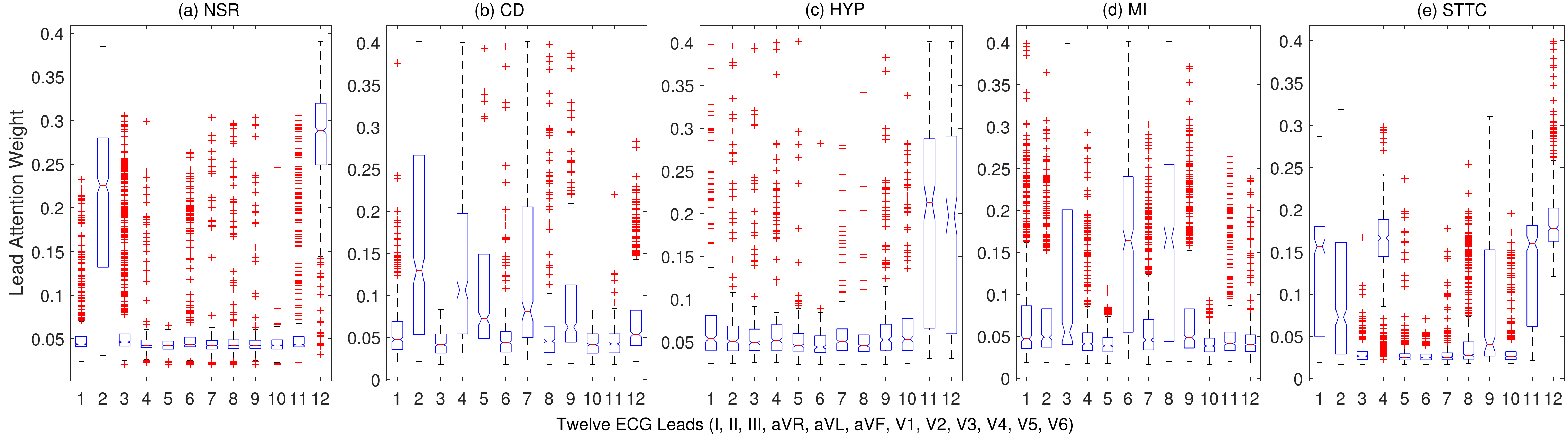}}
\caption{Boxplot of the spatial attention weights learned by the proposed ATCNN model for 12-lead ECG show the importance of each ECG-lead for diagnosing (a) NSR, (b) CD, (c) HYP, (d) MI and (e) STTC on the validation dataset.}
\label{BD04}
\end{figure*}

\begin{figure}[ht!]
\centerline{\includegraphics[height=1.6in,width=3.3in]{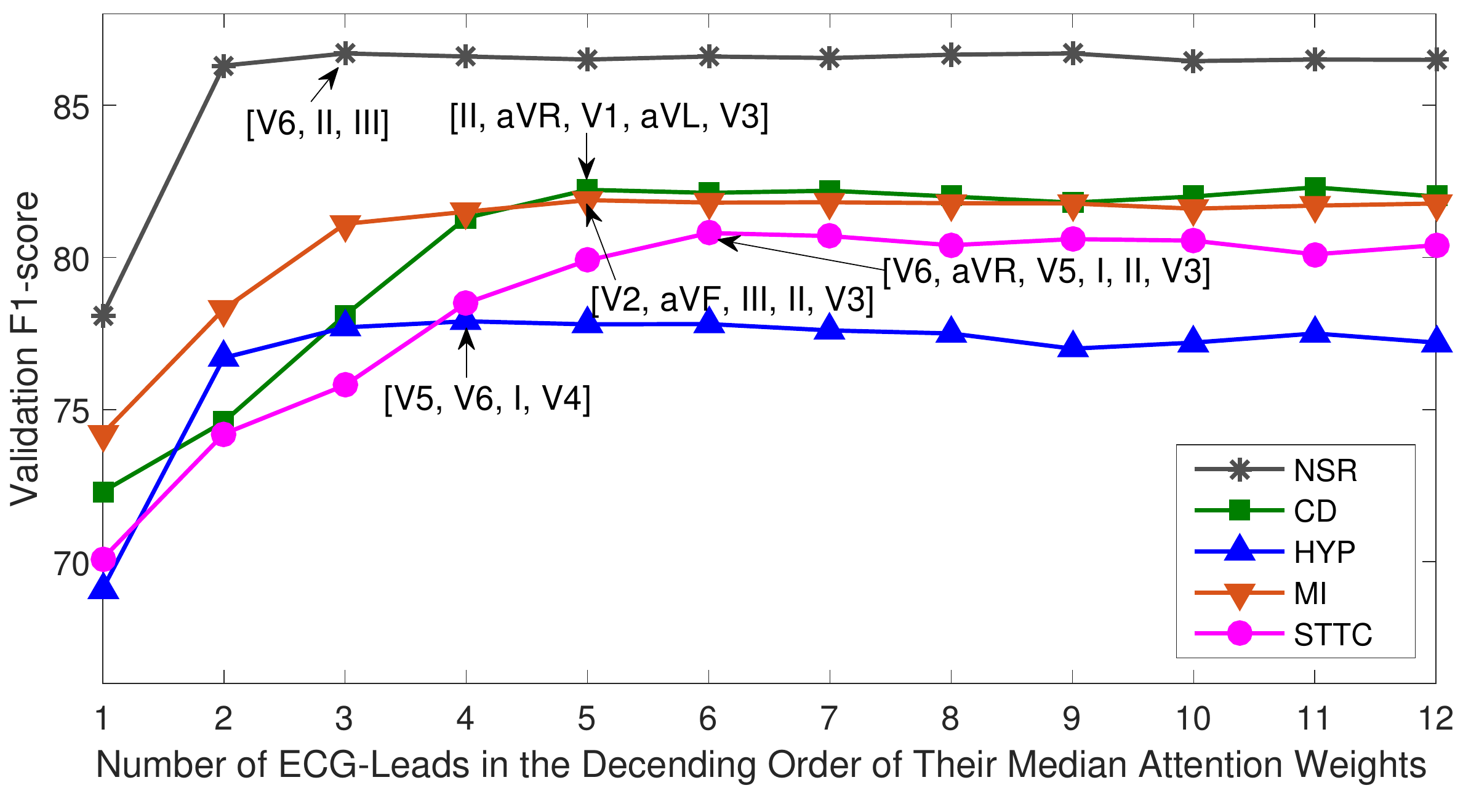}}
\caption{Optimal ECG-leads selection for each cardiac condition using the spatial attention weights learned by our model on the validation dataset.}
\label{BD05}
\end{figure}

\section{Discussion}
Automated interpretation of 12-lead ECG for identifying patients with multiple co-occurring disorders (multi-label) helps optimize patient-centered care delivery and reduce associated mortality. In this study, we demonstrated a problem transformation-based approach for automated multi-label ECG classification using an end-to-end ATCNN model. The proposed generic ATCNN model architecture can be applied to multiple cardiac disorders with high detection accuracy and can detect co-occurring disorders appearing in the same record. To the best of our knowledge, we are the first to explore the temporal convolutional layers with different receptive fields for the multi-label ECG classification. 

The proposed model is developed and evaluated on a large PTBXL-2020 multi-label ECG dataset. The performance is evaluated in terms of class-wise and average binary classification measures. In addition, an exact match measure is employed to evaluate the multi-label ECG classification performance. The experimental results of the proposed model (Table \ref{tab:results01})  demonstrated good detection performance for each cardiac condition. An ensemble of single-label binary classifiers with a fixed threshold of 0.5 on their predictions yielded promising results for multi-label ECG classification in terms of exact-match measure (Table \ref{tab:results01}). In practice, grouping patients with a single cardiac disorder and multiple co-occurring disorders plays a critical role in timely initiating life-saving treatment plans for high-risk multimorbidity patients. Thus, the decisions from the proposed framework are further analyzed to identify single and multimorbidity patients (Fig. \ref{BD01}(b)). The method correctly identified 408 single cardiac disorder patients out of 736 and 177 multimorbidity patients out of 465 from the test dataset. We report these results by matching true and predicted label vectors, ignoring the subset of correctly predicted labels.

\subsection{Ablation Study}
To verify the effectiveness of two crucial components of the proposed model, i.e., the TCNN layers and the temporal- and spatial-attention layers, we have performed the following experiments. First, the dilated and causal convolutional filters of the ATCNN are replaced with traditional convolutions while keeping attention layers as it is, and this modified network is retrained to perform the classification. Next, the temporal- and spatial-attention layers are removed from the ATCNN model (Fig. \ref{BD01}(a)), i.e., at each ECG-lead, the output feature maps of TCNN layers fed to the global average-pooling (GAP) layer, and these layer outputs of all the ECG-leads are concatenated and fed to the output layer to perform the classification. The experiments reveal that the proposed ATCNN model with dilated and causal convolutions outperforms the model with traditional convolutions with an improvement of nearly 3.5$\%$ in F1-score. Similarly, the ATCNN with temporal- and spatial-attention layers outperforms the model without attention layers with an improvement of nearly 5.7$\%$ in F1-score. These experiments verify that the combination of TCNN and attention layers effectively learn the 12-lead ECG's temporal dynamics and reduces its diagnostic redundancy by emphasizing clinically relevant information to improve classification performance.

 \begin{table}[t!] \scriptsize
 \caption{Comparison of F1-score for each cardiac condition on the test dataset using 12-lead ECG and optimal ECG-lead subset} 
 \centering 
 \setlength{\tabcolsep}{4pt}
\renewcommand{\arraystretch}{1.1}
 \begin{tabular}{c|c|c} \hline \hline
  \textbf{Class} & \textbf{F1-score Using 12-lead ECG} & \textbf{F1-score Using Optimal-Leads} \\ \hline
  NSR   &  86.01   &  85.86  \\
  CD   &  77.18   &  77.21  \\
  HYP  &  66.57   &  66.51  \\
  MI   &  76.41   &  76.38   \\
  STTC   &  76.34   & 76.22  \\ \hline
  \textbf{Average}   &  \textbf{76.51}   & \textbf{ 76.44}  \\ \hline        
 \end{tabular}
\label{tab:results02}
\end{table}

\begin{figure*}[ht!]
\centerline{\includegraphics[height=1.9in,width=6in]{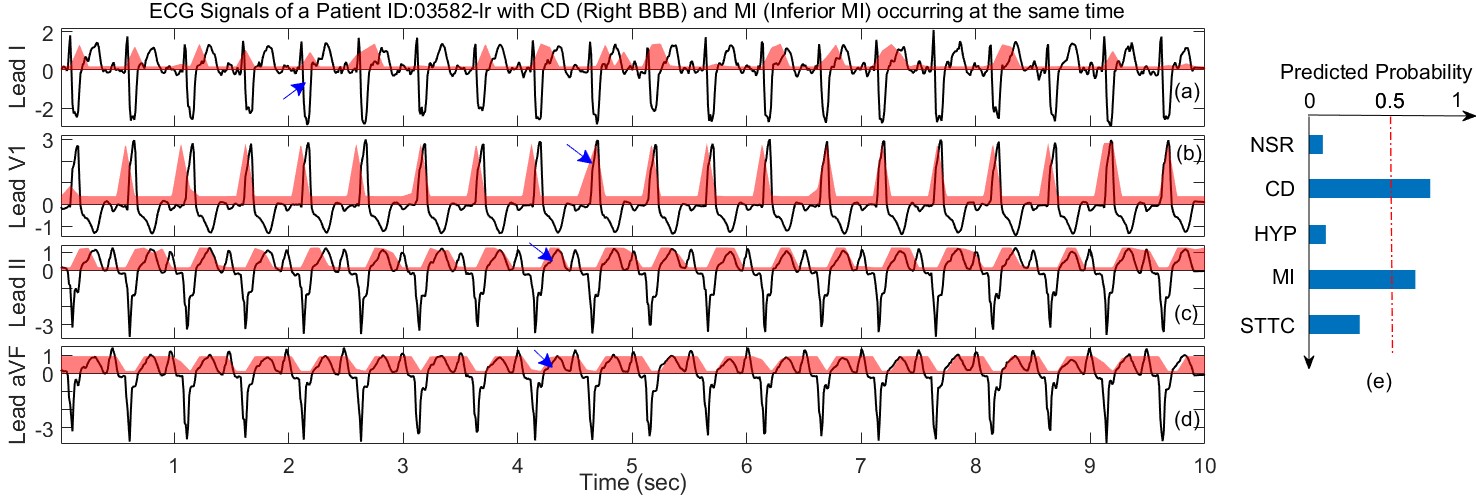}}
\caption{Visualization of temporal attention weights (red area height) assigned for different segments in optimal ECG-leads of a patient ID:03582-lr with right BBB and inferior MI occurring at the same time (a typical example from the test dataset). (a)-(b) Attention maps generated by the CD class binary classifier for ECG-leads I and V1. (c)-(d) Attention maps generated by the MI class binary classifier for ECG-leads II and aVF. As highlighted by the blue arrow, the red area height is more for the pathological ECG features to some extent. (e) Thus, the proposed approach correctly identifies the co-occurring cardiac disorders using a simple threshold of 0.5 on each class prediction probability.}
\label{BD06}
\end{figure*}

\subsection{Disease-Specific Optimal ECG-Leads Selection}
Fig. \ref{BD04} shows the distribution of spatial attention weights for the 12-lead ECG records of each cardiac condition. As can be seen, the proposed model provides different weights of importance for each ECG-lead based on their diagnostic relevance or redundancy in identifying the specific cardiac condition. Fig. \ref{BD05} illustrates the effectiveness of spatial attention weights in identifying optimal ECG-leads for each cardiac condition. Using the proposed ATCNN (trained on respective lead combination), the F1-score is computed for each class at different ECG-lead combinations arranged in the descending order of their median attention weights. For evaluating the single-lead ECG performance, the ATCNN architecture is modified as TCNN layers$\rightarrow$temporal attention layer$\rightarrow$output classification layer. The ECG-leads subset where the performance does not improve represents the optimal leads. As seen from Fig. \ref{BD04} and Fig. \ref{BD05}, for CD class leads, II, aVR, V1, aVL, and V3 contribute more to the classification. Similarly, ECG-leads V5, V6, I, and V4 for HYP; V2, aVF, III, II, and V3 for MI; and V6, aVR, V5, I, II, and V3 for STTC class. The reason for giving more weight to these ECG-leads is that clinicians often examine the abnormal QRS-morphology in leads V1 and aVL to diagnose CDs such as right BBB. The large R-wave in lateral leads V5, V6, and I helps detect HYP disease. Inferior leads II, III, and aVF and anterior leads V3 and V2 are used in the clinic to detect the ST-elevation inferior and anterior MI, respectively; the ST-depression and T-changes in lateral leads V5, V6, and I used in clinics to identify the lateral wall ischemia. Although ECG-lead aVR is often ignored in clinics, it helps diagnose CD and STTC patients \cite{Gorgels2001}. Moreover,  the similar performance for 12-lead ECG and optimal ECG-leads (Table \ref{tab:results02}) demonstrate that the proposed model automatically identifies the diagnostically relevant and redundant ECG-leads for improving classification performance.

\subsection{Model Interpretability}
Fig. \ref{BD06}(a)-(d) shows the model interpretability visualization in terms of temporal attention weight maps for a key ECG-leads of a patient with complete RBBB and inferior MI occurring at the same time. The proposed approach correctly identified these co-occurring disorders using simple thresholding on the prediction probability of single-label binary classifiers (Fig. \ref{BD06}(e)). Thus, the attention weight maps for the optimal ECG-leads of RBBB (Fig. \ref{BD06}(a)-(b)) and inferior MI (Fig. \ref{BD06}(c)-(d)) disorders are extracted from their corresponding binary classifier. As can be seen, the temporal attention layer provides different weights for different ECG features based on their relevance in the diagnosis process. In practice, clinicians examine (i) the deep S-waves in the lead I and broad QRS-complexes in the lead V1 for diagnosing RBBB disorder and (ii) the distinct ST-segment elevations in the inferior leads II and aVF for diagnosing inferior MI. Interestingly, the temporal attention maps learned by the model indeed provide more weight to these pathological ECG features (Fig. \ref{BD06}(a)-(b) and Fig. \ref{BD06}(c)-(d)) during the diagnosis process. In summary, the temporal attention maps generated by the proposed model match with the clinical evidence to some extent, which aided in improved performance (Fig. \ref{BD06}(e)) and model interpretability.

\begin{table}[t!] \scriptsize
\caption{Comparison of the proposed method with existing multi-label ECG classification methods} 
\centering 
\setlength{\tabcolsep}{2.3pt}
\renewcommand{\arraystretch}{1.2}
\begin{tabular}{p{0.1\linewidth} p{0.35\linewidth} p{0.28\linewidth} p{0.22\linewidth}} \hline \hline
 \textbf{Ref.}  & \textbf{Method} & \textbf{Database}  &  \textbf{Performance} \\ \hline
 
 \cite{Li2021}    & \begin{itemize} 
\vspace*{-\baselineskip}
\item 117 handcrafted features
\item Feature selection
\item Multi-objective model \vspace*{-\baselineskip}
\end{itemize}   & \begin{itemize} 
\vspace*{-\baselineskip}
\item CPSC (n = 6,877)
\item $\#$multi-label 12-lead ECG records = 476 \vspace*{-\baselineskip}
\end{itemize}   &  F1-score = 60.8$\%$  \\ \hline

 \cite{Cheng2020}    & \begin{itemize} 
\vspace*{-\baselineskip}
\item Extraction of single-lead compressed ECG signals
\item Residual CNN model \vspace*{-\baselineskip}
\end{itemize}   & \begin{itemize} 
\vspace*{-\baselineskip}
\item MITDB (n = 47)
\item $\#$multi-label 1-lead ECG records = 41 \vspace*{-\baselineskip}
\end{itemize}   &  F1-score = 98.4$\%$  \\ \hline 
 
\cite{Jia2019}   & \begin{itemize} 
\vspace*{-\baselineskip}
\item Sequence generation CNN module
\item Multi-task CNN module
\item Voting ensemble strategy \vspace*{-\baselineskip}
\end{itemize}   & \begin{itemize} 
\vspace*{-\baselineskip}
\item CPSC (n = 6,877)
\item $\#$multi-label 12-lead ECG records = 476 \vspace*{-\baselineskip}
\end{itemize}   &  F1-score = 87.2$\%$  \\ \hline

 \cite{Ganeshkumar2021}   & \begin{itemize} 
\vspace*{-\baselineskip}
\item 12-lead ECG beat matrix
\item DCNN-based model
\item Model interpretability \vspace*{-\baselineskip}
\end{itemize}   & \begin{itemize} 
\vspace*{-\baselineskip}
\item CPSC (n = 6,877)
\item $\#$multi-label 12-lead ECG records = 476 \vspace*{-\baselineskip}
\end{itemize}   &  F1-score = 96.7$\%$  \\ \hline

\cite{Jin2021}    & \begin{itemize} 
\vspace*{-\baselineskip}
\item Single-lead ECG extraction
\item CNN+BiLSTM+beat-level attention+rhythm-level attention model
\item Model interpretability \vspace*{-\baselineskip}
\end{itemize}   & \begin{itemize} 
\vspace*{-\baselineskip}
\item MITDB (n = 47)
\item CPSC (n = 6,877)
\item $\#$multi-label 1-lead ECG records = 517 \vspace*{-\baselineskip}
\end{itemize}   &  F1-score (MITDB) = 80.5$\%$; F1-score (CPSC) = 60.1$\%$\\ \hline

 \cite{Yoo2021}    & \begin{itemize} 
\vspace*{-\baselineskip}
\item Residual CNN model with squeeze and excite blocks
\item Soft voting strategy using k-labelset \vspace*{-\baselineskip}
\end{itemize}   & \begin{itemize} 
\vspace*{-\baselineskip}
\item PTBXL (n = 21,836)
\item $\#$multi-label 12-lead ECG records = 4,798 \vspace*{-\baselineskip}
\end{itemize}   &  Exact match measure = 61.2$\%$  \\ \hline

 \textbf{Proposed}    & \begin{itemize} 
\vspace*{-\baselineskip}
\item Temporal CNN layers with different dilation rates followed by temporal and spatial attention layers
\item Optimal ECG-leads selection for each disorder
\item Model interpretability \vspace*{-\baselineskip}
\end{itemize}   & \begin{itemize} 
\vspace*{-\baselineskip}
\item PTBXL (n = 21,836)
\item $\#$multi-label 12-lead ECG records = \textbf{4,798} \vspace*{-\baselineskip}
\end{itemize}   &  \textbf{F1-score = 76.5$\%$}; \,  \textbf{Exact match measure = 66.4$\%$} \\ \hline       
\end{tabular}
\label{tab:Comparison}
\end{table}

\subsection{Comparison with Existing Methods}
Table \ref{tab:Comparison} compares the performance of the proposed ATCNN model with existing multi-label ECG classification methods. Existing methods \cite{Li2021,Sun2020,Cheng2020,Jia2019,Yoo2021,Ganeshkumar2021,Jin2021} mostly used algorithm adaptation approaches for multi-label ECG analysis. However, these methods require complex models with rigorous threshold adjustment to achieve good multi-label ECG classification performance. In contrast, the proposed method is based on the problem transformation approach, where the complex problem is transformed into a set of simple single-label binary classifiers. An ensemble of predictions from these classifiers provided promising performance for multi-label ECG classification (Table \ref{tab:Comparison}). Methods \cite{Li2021,Sun2020,Ganeshkumar2021,Jin2021} employed a small CPSC dataset containing only 476 multi-label 12-lead ECG records for the evaluation. Thus, the higher F1-score of these methods may be attributed to the correct detection of single-label disorders. The proposed method is validated on a large PTBXL dataset consisting of 4,798 multi-label ECG records.
Moreover, the dataset contains ECG records of patients with acute disorders such as MI, HYP, and CD; thus, detecting co-occurrence of these disorders helps clinicians identify high-risk patients and initiate timely treatment. The methods \cite{Cheng2020,Jin2021} based on single-lead ECG signals may not be suitable for detecting localized disorders such as MI and HYP. Compared to the method \cite{Yoo2021}, the proposed model provides an improved exact match measure on the same PTBXL dataset. Similar to methods \cite{Ganeshkumar2021,Jin2021}, the proposed DL-based model also shows promising model interpretability results; thus, it may help clinicians and researchers to understand the pathological ECG features that led to the diagnosis. To the best of our knowledge, we are the first to identify the disease-specific optimal ECG-leads using the spatial attention weights (Table \ref{tab:Comparison}). Finally, compared to the existing methods, incorporating TCNN layers with different dilation rates followed by attention layers in the proposed model effectively learned the multi-scaled pathological ECG features dynamics, improving performance. In order to generate the multi-label classification decision for the 12-lead ECG input, the proposed framework takes an acceptable average test run-time of 9.59 ms.

\section{Conclusion and future directions}
This paper presents a problem transformation-based approach named ATCNN for multi-label ECG classification using 12-lead ECG records. Specifically, a set of single-label binary classifiers, one for each cardiac disorder, is designed, and the ensemble of predictions from these classifiers generated improved multi-label ECG classification performance. The ATCNN model with a stack of TCNN layers that operate at different dilation rates followed by temporal and spatial attention layers effectively learned the multi-scaled pathological ECG features dynamics. The experimental results demonstrated that our approach is good at classifying multiple co-occurring cardiac disorders appearing in the same ECG record. Moreover, the analysis of spatial attention weights showed that the proposed model automatically identifies the optimal ECG-lead subset and reduces the diagnostic redundancy of the 12-lead ECG. We also demonstrated the advantage of our approach for identifying the single and multimorbidity patients with encouraging model interpretability results; thus, the method has the potential to screen patients at high-risk and help clinicians initiate timely treatment.

There are some limitations of this study that need to be addressed
in future works. This study employed the problem transformation-based approach; thus, it does not consider the label correlation information; still, the method demonstrated its effectiveness for multi-label ECG analysis on the current dataset. Moreover, using ECG alone to detect multimorbidity patients may have systematic limitations; thus, combining ECG features with the other multi-modal information such as patients’ risk factors and echocardiogram findings may improve the performance further.

\end{document}